# *In silico* Prediction of Mozenavir as potential drug for SARS-CoV-2 infection via Binding Multiple Drug Targets

Estari Mamidala[a*], Rakesh Davella[b], Swapna Gurrapu[c], Munipally Praveen Kumar[d], Abhiav[e]

[a,b,c,d] Infectious Diseases Research Lab, Department of Zoology, Kakatiya University, Warangal, Telangana-506 009, India

[e] Division of ISRM, Indian Council of Medical Research (ICMR), Department of Health Research, New Delhi, India

*Corresponding should be addressed to Estari Mamidala: **drestari@kakatiya.ac.in**

**ABSTRACT**

Since the epidemic began in November 2019, no viable medicine against SARS-CoV-2 has been discovered. The typical medication discovery strategy requires several years of rigorous research and development as well as a significant financial commitment, which is not feasible in the face of the current epidemic. Through molecular docking and dynamic simulation studies, we used the FDA-approved drug mezonavir against the most important viral targets, including spike (S) glycoprotein, Transmembrane serine protease 2 (TMPRSS2), RNA-dependent RNA polymerase (RdRp), Main protease (Mpro), human angiotensin-converting enzyme 2 (ACE-2), and furin. These targets are critical for viral replication and infection propagation because they play a key role in replication/transcription and host cell recognition. Molecular docking revealed that the antiviral medication mozenavir showed a stronger affinity for SARS-CoV-2 target proteins than reference medicines in this investigation. We discovered that mozenavir increases the complex's stability and validates the molecular docking findings using molecular dynamics modelling. Furin, a target protein of COVID-19, has a greater binding affinity (-12.04 kcal/mol) than other COVID-19 target proteins, forming different hydrogen bonds and polar and hydrophobic interactions, suggesting that it might be used as an antiviral treatment against SARS-CoV-2. Overall, the present *in silico* results will be valuable in identifying crucial targets for subsequent experimental investigations that might help combat COVID-19 by blocking the protease furin's proteolytic activity.

# INTRODUCTION

In the Hubei province in central China, a novel virus belonging to the human coronavirus family was found and called SRAS-CoV-2[1]. Although the source of zoonotic infection has yet to be determined, this new virus is most likely the product of a recombination between bat coronavirus and pangolin coronavirus. SRAS-CoV-2 causes a severe respiratory illness in humans with a significant mortality rate[2]. According to the WHO Coronavirus Disease 2021 Situation Report, as of June 9, 2021, up to 173,613,778 patients worldwide had tested positive for COVID-19, with 3,742,856 deaths[3], and the Ministry of Health and Family Welfare (MOHFW), Government of India report, active cases are 123,141,5 with 3,53,528 deaths[4]. COVID-19 has grown swiftly in over 212 nations[5] after its initial appearance, and the World Health Organization labelled it a global health emergency on March 11, 2020. There is presently no particular therapy for the condition, and extensive research is ongoing to find vaccinations and medications to combat it.

SARS-CoV-2 is a member of the CoV genus in the Coronavirinae subfamily of the Nidoviral order[6]. It most closely resembles SARS-CoV-1 (2002–03), the coronavirus strains that caused the two most recent epidemics, in 2002–2003 and 2012, respectively, with 79.5 percent sequence identity and the same mechanism for host cell entry, via the furin and angiotensin-converting enzyme-2 (ACE-2) surface proteins of the cell[7]. Researchers are concentrating on "drug repositioning" due to the urgent need for effective treatments to slow the disease's progression and the lengthy process of developing new medications. The majority of the medications licenced or developed for other reasons that have been suggested for the treatment of SARS-CoV-2 and are now being tested in clinical trials are enzyme inhibitors that target the virus's RNA-dependent RNA polymerase (RDRP) or protease. RDRP inhibitors include favipinavir, ribavirin, remdesivir, and galidesivir, whereas viral protease inhibitors include lopinavir, ritonavir[8], and danoprevir. In addition, medications such as chloroquine, hydroxychloroquine[9], and APN01[10] have been suggested to block viral particle entry and development.

The three-dimensional structures of SARS-CoV-2 SARS in the receptor-binding region led to the development of numerous ways to identify a possible target for a therapeutic candidate to be utilised in the battle against SARS-CoV-2, including furin, ACE2, TMPRSS2, Spike protein,

Main protease, and RdRp. The S1 domain of the Coronavirus Spike protein is responsible for receptor binding, whereas the S2 domain mediates membrane fusion in the early stages of viral infection[11]. The process of SARS-CoV infecting the host necessitates essential cleaving processes, which alter SARS-infectious CoV's ability. The spike protein is broken into S1 and S2 subunits by type-II transmembrane serine protease (TMPRSS2)[12]. The first component attaches to a receptor on the surface of the host cell, while the second is in charge of fusing the viral envelope to the cell membrane. Furin is a kind of proprotein convertase (PC) that may cleave precursor proteins with certain patterns to create biologically active mature proteins. Furin-like enzymes cleave the viral envelope glycoprotein into a functional binding virus receptor and a fusogenic transmembrane protein, which is essential for virus cell entrance and infectivity[13]. So yet, no small molecule inhibitor of furin has been discovered that has both excellent effects and excellent specificity. RNA-dependent RNA polymerase (RdRp; nsp12) is also required for RNA replication and transcription[14].

Despite substantial research on SARS-CoV-2, no licenced treatment against the virus has yet to be developed. As a result, sensitive and efficient antiviral medicines against zoonotic coronaviruses, such as SARS-CoV-2, are urgently needed. Because it takes a long time for novel antiviral medications to be licenced, several studies have been done to assess the effectiveness of existing licenced medications against SARS-CoV-2. Several medicines have been discovered to have antiviral effect against SARS-CoV-2. Old antimalarials (chloroquine phosphate, chloroquine, and hydroxychloroquine)[15], an anthelmintic (ivermectin)[16], viral RNA polymerase inhibitors (remdesivir and favipiravir)[17], and viral protease inhibitors (remdesivir and favipiravir)[18] are among them. Mozenavir (DMP-450) is an antiviral medication that was created as an HIV/AIDS therapy. It works as an inhibitor of HIV proteases and binds to the target with a high affinity that was chosen for this investigation[19,20]. In order to investigate the binding affinity of various viral proteins to mozenavir with possible antiviral activity against SARS-CoV-2, the current work used molecular docking and dynamic simulations.

**RESULTS & DISCUSSION**

Six SARS-CoV-2 targeting proteins, including spike glycoprotein, TMPRSS2, RdRp, Mpro, human ACE-2, and furin, were docked with the antiviral medication mozenavir in order to investigate a possible treatment target. Because hydrogen bonds (H-bonds) are important for ligand-target protein interactions and binding affinity, we investigated the binding affinity of mozenavir among SARS-CoV-2 target proteins using H-bond interactions (Table 1). The results of the molecular docking studies are compared to previously reported drug candidates (Decanoyl-RVKR-chloromethylketone, Arbidol, Camostat, Remdesivir, Indinavir, and Chloroquine phosphate) that inhibit the furin, Spike glycoprotein, TMPRSS2, RdRp, mpro, and ACE-2 target proteins of SARS-CoV-2) that inhibit the furin (Table 2).

The virtual screening of mozenavir against the SARS-CoV-2 target proteins using molecular docking demonstrated a significant interaction with greater docking energy and binding affinities, as shown in figure 3. The antiviral medication mozenavir docked with COVID-19's therapeutic targets in the active region of the protein. Table-1 shows the binding energy values for all COVID-19 targets, which vary between (7.08) and -12.04 kcal/mol. Furin, the COVID-19 target protein, had higher binding affinities (lowest binding energy) (-12.04 kcal/mol), which were higher than the binding affinities of the other COVID-19 target proteins. According to the findings of the molecular docking investigations, mozenavir was discovered to be a candidate that inhibits all of the COVID-19 target proteins.

Discovery Studio Visualizer and UCSF Chimera were used to show and assess mozenavir docking with all protein targets binding affinities. Figures 4 to 9 depict the surface structure, 2D, and 3D interactions of mozenavir with all target sites.

Based on binding energy acquired through molecular docking against chosen COVID-19 targets, Furin (-12.04 kcal/mol), ACE-2 (-9.71 kcal/mol), Mpro (-8.79 kcal/mol), RdRp (-7.32 kcal/mol), Spike glycoprotein (-7.09 kcal/mol), and TMPRSS2 (-7.08 kcal/mol) with chosen antiviral medication mozenavir, it is predicted that, the furin is to be the best target for COVID-19 infection.

**Docking of the mozenavir with the Furin**

The SARS-CoV-2 spike protein features a furin cleavage site that was not seen in any other betacoronavirus subtype B. Based on our findings, we believe that SARS-greater CoV-2's infectious nature than other coronaviruses is due to the existence of a redundant furin cut site in its Spike protein, which leads to increased membrane fusion efficiency[24]. As a result, the SARS-CoV-2 may take advantage of the high expression of furin and other furin-like enzymes present in human lung, liver, and brain tissues[25] to activate S protein, resulting in increased infection, virulence, and viral transmission. Interfering with the cleavage of the SARS-CoV-2 S protein processing by compounds might be a promising antiviral strategy. As a reference chemical, decanoyl-RVKR-chloromethylketone, a well-known medication that has previously been described as a furin inhibitor, was investigated. Tables 1 and 2 show the docking score of mozenavir in comparison to the reference chemical. We employed bioinformatics methods to identify licenced and experimental drugs (Mozenavir) as probable furin inhibitors in the present study.

Figure 2 depicts docking studies of the mozenavir-furin interaction. Furin with mozenavir had the lowest binding energy of -12.05 kcal/mol. Furin discovered that the binding energy of the reference medication decanoyl-RVKR-chloromethylketone is -6.89 kcal/mol, which is greater than the binding affinity of the test medication mozenavir. Despite the fact that Furin-mozenavir has the most hydrogen bonds (five) (fig 4, green coloured). With mozenavir, Tyr308, Gly265, Gly255, Asp154, and Val231 showed a strong hydrogen interaction. In the development of a hydrophobic contact with mozenavir, the amino acid residues Glu236, Pro256, Trp254, Leu227, His194, Gly229, Asp264, Asp191, Glu230, and Gly229 were implicated. As a result, mozenavir might be regarded a possible furin inhibitor.

**Docking of the mozenavir with the ACE-2**

We reasoned that blocking viral entrance into the host cell would limit viral burden. One of the enzymes on the host cell surface that allows coronavirus to enter the host is ACE2. As a result, we chose this receptor protein for our research. The SARS-CoV-2 spike protein (S) binds to ACE2, which may aid in the transfer of COVID-19 from humans to humans[26]. ACE2 also aids endocytosis and endosome-containing coronaviruses. Targeting as a viral entry inhibitor is a unique anti-SARS-CoV-2 therapeutic strategy. As a result, we looked at the FDA-approved medication mozenavir for suppressing human ACE2. As a reference chemical, chloroquine phosphate, a well-known medication that has previously been claimed to be an ACE2 inhibitor, was investigated. Table 2 shows the docking score of mozenavir in comparison to the reference molecule.

When compared to other target proteins such as Mpro, Spike glycoprotein, and RdRp, as well as the reference medication Chloroquine phosphate (-7.88 kcal/mol), docking data indicated that mozenavir had the greatest binding affinity to the active site of the ACE-2 protein (-9.71 kcal/mol). It also established four H-bonds with Arg393, Tyr385, His378 and Glu375 amino acid residues found in the protein's putative active site. ACE-2's Asn394, Asp350, His401, Ala348, His374, Pro346, His345, Phe504, Thr347, and Phe40 amino acid residues reacted hydrophobically with mozenavir. When these data are analysed, it is clear that the antiviral medicine mozenavir has a low binding energy with ACE-2, indicating that it should be investigated further in vitro and in vivo before being considered as a possible COVID-19 medication.

**Docking of the mozenavir with the SARS-CoV-2 Mpro**

SARS-CoV-2 Mpro, also known as 3C-like proteins, is a cysteine protease with three domains (domains I-III) and a 33.8 kDa size[27]. Mpro is implicated in polyprotein cleavage at eleven conserved sites, resulting in mature and intermediate non-structural proteins[27]. SARS-CoV-2 Mpro has a noncanonical dyad of Cys145-His41 between domains I and II, which is related to domain III through a loop[28]. The amino acids Cys145 and His41 are important for substrate recognition[29]. As a result, this protein was chosen as a potential COVID-19 therapeutic target, and intermolecular interactions, including binding energy, were tallied in table 1 and shown in figure 6. Indinavir was used as a reference molecule since it is a well-known medicine that has previously been linked to Mpro inhibitors. Table 2 shows the docking score of mozenavir in comparison to the reference molecule.

When compared to the reference medication Indinavir (-7.11 kcal/mol), docking studies indicated that mozenavir had the greatest binding affinity (lowest binding energy) to the active site of the protein with - 8.79 kcal/mol against Mpro. Furthermore, mozenavir established two H-bonds with Asp153 and Asp245 amino acid residues found in the protein's predicted active site, as well

as hydrophobic interactions with Ile249, Phe294, His246, Gly109, Ile200, Gln110, Thr292, Asn203, Val202, Thr111, and Asn151.

**Docking of the mozenavir with the SARS-CoV-2 RdRp**

RdRp, also known as nsp12, is made up of two extra subunits, nsp7 and nsp8[30]. The N-terminal nido-virus RdRp-associated nucleotidyl transferase (NiRAN) domain, an interface domain, and a C-terminal end domain make up RdRp's structural structure[31]. RdRp's primary role is to catalyse viral replication beginning at the 3′-poly-A end. RdRp catalyses the replication of the RNA genome by using the (+) RNA strand as a template to create a complementary () RNA strand[32]. RdRp's active site is made up of -helices, an antiparallel -strand, and catalytic aspartate[33]. RdRp is an important antiviral therapeutic target because of its function in the replication cycle of coronaviruses[34]. Remdesivir was used as a reference chemical since it is a well-known medicine that has previously been identified as a RdRp inhibitor.

In comparison to the reference drug remdesivir (-4.7 kcal/mol), we conducted molecular docking tests of the antiviral medicine mozenavir against the RdRp protein of the SARS-CoV-2 and found that mozenavir had the lowest binding energy (–7.32 kcal/mol) and the greatest binding affinity (–7.32 kcal/mol). Mozenavir binds to the active site of RdRp with -7.32 kcal/mol and forms several interactions with residues such as Tyr515, Trp509, Ala375, Leu371, Met380, Tyr374, Phe340, Phe368, Phe506, and Leu372 without creating any hydrogen bonds, according to the current research (fig 7). Based on these findings, the antiviral medication mozenavir might be a promising RdRp inhibitor in the fight against SARS-CoV-2.

**Docking of the mozenavir with the SARS-CoV-2 splike glycoprotein**

Virtual screening aided molecular docking was used on the binding pocket of spike proteins to examine a possible antiviral medication targeting the spike protein of SARS-CoV-2. The viral entrance into the host's cellular system with the help of the ACE-2 receptor has been thoroughly documented[35]. Glycosylated spike 1 (S1) interacts to the ACE-2 receptor on the surface of the human host cell and mediates viral entry[36]. Table 1 shows the docking scores of mozenavir, which were chosen for the research of Spro inhibition in SARS-CoV-2. Arbidol was used as a reference drug since it is a well-known medicine that has been shown to increase glycoprotein inhibitors in the past.

Mozenavir has a docking score of -7.09 kcal/mol with spike glycoprotein. Mozenavir, on the other hand, has the lowest binding affinity for SARS-CoV-2 spike proteins when compared to other protein targets such as furin, ACE-2, and RdRp. However, this target protein generated two hydrogen bonds with Asp415 and Ser370, as well as electrostatic interactions with Val369, Gly368, Tyr367, Asp414, Cys366, Lys365, Pro399, and Gln401 (ig 8). When these data are analysed, it can be shown that mozenavir has a relatively low binding energy with Spro of the SARS-CoV-2, indicating that further in vitro and in vivo research is needed before they can be considered as prospective COVID-19 medicines.

**Docking of the mozenavir with TMPRSS2**

The spike glycoprotein's S1 subunit binds to the cellular receptor first, followed by ACE-2 priming S protein by the host transmembrane serine protease 2 (TMPRSS2), which breaks out the viral S protein right upstream of the fusion peptide, aiding membrane fusion through irreversible conformational changes[37]. Inhibition of TMPRSS2 may assist to prevent the SARS-CoV-2 virus from infecting human cells. A reduction in TMPRSS2 expression and activity has also been shown in previous studies to be a safe and effective technique for treating viral infection caused by viruses[38,39]. As a result, several researchers are working to find an effective drug to suppress TMPRSS2, which has been chosen as a coronavirus target protein. Camostat was used as a reference molecule since it is a well-known medication that has previously been described as a TMPRSS2 inhibitor.

When compared to the reference molecule camostat (-5.9 kcal/mol), docking studies indicated that mozenavir had the greatest binding affinity and lowest binding energy (-7.08 kcal/mol) to the active site of the TMPRSS2 protein. Mozenavir also formed one H-bond with Tyr337, an amino acid residue found in the TMPRSS2 protein's predicted active site, as well as ten hydrophobic interactions with Ala490, Gln487, Pro335, His334, Ile332, Ser333, Asn303, Val298, Val331, and Lys330, making it a promising TMPRSS2 inhibitor candidate (Figure-9).

**Molecular dynamics simulation**

According to a molecular docking investigation of mozenavir with several COVID-19 targets, mozenavir had the greatest binding affinity for furin. As a result, a molecular dynamics simulation was used to assess the stability of the target protein furin alone and in combination with mozenavir.

**Estimation of root mean square deviations (RMSDs)**

Figure 4A shows root mean square deviations (RMSDs) in the backbone of furin alone or in combination with mozenavir as a function of simulation duration as compared to the start frame. Due to the equilibration of the initial protein structure, substantial fluctuations in RMSD values (up to 0.3 nm or 3.0) of protein alone were seen over the first 2000 ps (2 ns). As a result, the simulation period was increased from 2000 to 6000 ps, and the RMSD values were altered and exceeded the allowed limit of 0.2 nm. Throughout the simulation duration, the RMSD values of furin with bound ligand mozenavir were under the top limit of 0.2 nm (2 Ao) (1000 ps). As a result of the creation of complementary connections between protein and ligands, a stable protein-ligand complex was formed, as shown by steady RMSD values.

**Root mean square fluctuations (RMSFs) determination:**

Furthermore, the root mean square fluctuations (RMSFs) along the furin side chains were monitored in order to track any conformational changes caused by mozenavir binding (fig 11). The data demonstrated that the binding site residues exhibited less variation when furin with 472 amino acids was complexed with mozenavir medication candidate. The average RMSF values for furin alone and complex with mozenavir were 0.21 and 0.23 nm, respectively (figs. 5b, 6b & 7b). As a consequence of the RMSFs, the creation of a stable protein-ligand complex was verified.

The closeness of the ligand mozenavir to active site residues of the target protein furin may imply greater binding, according to dynamic modelling studies. As shown in Figure 1, mozenavir was able to sustain a low ligand mobility root-mean square deviation (RMSD) of less than 0.3 nm (fig 11). This graph was created by superimposing the furin receptor on its reference structure. Finally, molecular dynamics study has shown that Mozenaivr may experience greater conformational changes over simulation time.

**CONCLUSION**

In conclusion**,** COVID-19 is a worldwide illness that has a significant fatality rate. The antiviral medication mozenavir was chosen to interact with various SARS-CoV-2 targets, including spike glycoprotein, TMPRSS2, RdRp, Mpro, human ACE-2, and furin, in this investigation. Our goal was to find a chemical that might bind several SARS-CoV-2 targets since viral drug targets are more prone to alterations. Mozenaivr was discovered to interact with various targets and to be particularly effective at blocking all six SARS-CoV-2 targets (spike glycoprotein, TMPRSS2, RdRp, Mpro, human ACE-2, and furin) with considerable binding affinities as compared to reference medicines. When compared to other COVID-19 targets, we chose six target proteins for molecular docking with mozenaivr, and the medication has the greatest binding affinity (lowest binding energy), which is -12.04 kcal/mol with furin. The RMSD of the furin-mozenavir complex in the upper lomit was 0.2 nm, suggesting that the mozenaivr underwent excellent conformational changes during binding and retained tight affinity with the furin's binding site. Together, we think that mozenavir has the potential to suppress SARS-CoV-2 by binding to various pharmacological targets, including furin, and that additional in vitro and in vivo research is warranted.

**MATERIALS AND METHODS**

**Protein Retrieval and Preparation:**

The 3D structures of the S glycoprotein (PDB ID = 2AJF), TMPRSS2 (PDB ID = 7MEQ), RdRp (PDB ID = 7B3C), Mpro (PDB ID = 6Y2E), human ACE-2 (PDB ID = 7DF4), and Furin (PDB ID = 5JXH) from the Research Collaboratory for Structural Bioinformatics (RCSB) Protein Data Bank[21]. All of the proteins found were employed in molecular docking experiments with the medication Mozenavir.

**Ligand Preparation:**

The 3D structure of Mozenavir (Figure 2), as well as arbidol, camostat, remdesivir, indinavir, chloroquine phosphate, and decanoyl-RVKR-chloromethylketone, which are all proposed drug candidates for treating COVID-19, were downloaded in.sdf format from the PubChem database (http://www.pubchem.ncbi.nlm.nih). Open babul software was used to transform their three-dimensional (3D) structures to PDB format, and all of the structures were energy reduced and converted to PDBQT format using AutoDock Tools.

**Molecular Docking:**

Computer-based approaches were employed to examine the structure of ligand-protein complexes and metabolic pathways in the docking investigation of possible therapeutic medication candidates employed against COVID-19 across the globe. AutoDock 4.2 (The Scripps Research Institute, La Jolla, CA, USA) was used for all docking tests since it has a faster running time due to multiple core processors and better accuracy for ligands with more than 20 rotatable bonds. Using AutoDock tools 1.5.6, the protein molecules and ligands were transformed to their correct readable file format (pdbqt). The blind docking approach was used in all of the docking studies, which included creating a grid box big enough to encompass the whole protein structure and include any potential protein-ligand interactions. Each docking operation had a total of ten runs. Furthermore, the maximum iterations were 2000, with a 100 Kcal/mol energy barrier. The default settings for all other programme settings were used. The lowest docked binding energy was used to choose the optimal conformations for each docking procedure. Discovery Studio Visualizer 2.5 (Accelrys Software Inc., San Diego, CA, USA) and UCSF-Chimera were used to create the final representation of the docked structure.

**Molecular Dynamic Simulation**

Molecular dynamics (MD) simulation for each of the anticipated receptor-ligand complexes was performed to confirm the correctness of the binding interaction findings. GROMACS 2018[22] was used to run all simulations, which used the AMBER 99SB-ILDN force field[23]. Energy minimization, equilibration (NVT and NPT), and MD simulation with 2 femtosecond integration steps for 10 ns were all followed as usual. The system was neutralised by introducing ions and then relaxing using the energy minimization method. MD simulations with an acceptable beginning velocity descend to a local minimum via the sharpest descent route on the potential energy surface. Prior to the 10ns production simulation, a 1ns temperature and pressure equilibrium step was done. g_rms, g_rmsf, and g_gyrate were used to determine the root mean square deviation (RMSD), root mean square luctuation (RMSF), and Radius of gyration (Rg). In addition, the generation of hydrogen bonds between protein-ligand and protein solvent was estimated. A protein alone or with ligand that could keep its RMSD under 1 nm was regarded fixed, less than 2 nm was termed stable, and more than 2 nm was termed non-stable.

**ACKNOWLEDGMENTS**

Authors are thankful to Head, Department of Zoology, Kakatiya University, Warangal for providing necessary research facilities and infrastructure.

**COMPETING FINANCIAL INTERESTS**

The authors declare no competing financial interests

**REFERENCES**

[1]  The, L. Emerging understandings of 2019-nCoV. Lancet 2020;395:311.

[2]  Li Q. Early Transmission Dynamics in Wuhan, China, of Novel Coronavirus- Infected Pneumonia. N Engl J Med 2020;382:1199-1207.

[3]  WHO. WHO Coronavirus Disease (COVID-19) Dashboard. Geneva: World Health Organization 2021.

[4]  MOHEW. https://www.mohfw.gov.in/. Published 2020. Accessed June 9, 2021.

[5]  Bedford J, Enria D, Giesecke J, Heymann DL, Ihekweazu C, Kobinger G. WHO Strategic and Technical Advisory Group for Infectious Hazards COVID-19: Towards controlling of a pandemic. Lancet 2020;395:1015–1018.

[6]  Chan JFW, Kok KH, Zhu Z, Chu H, To KKW, Yuan S, Yuen KY. Genomic characterization of the 2019 novel human pathogenic coronavirus isolated from a patient with atypical pneumonia after visiting Wuhan. Emerg. Microbes Infect 2020;9:221–236.

[7]  Zhou P, Yang XL, Wang XG, Hu B, Zhang L, Zhang W. A pneumonia outbreak associated with a new coronavirus of probable bat origin. Nature 2020;579:270–273.

[8]  Mamidala. An In silico approach for identification of inhibitors as a potential therapeutics targeting SARS-Cov-2 protease. Asian J Pharmaceut Res Health Care 2020;12:3-9.

[9]  Mamidala E, Davella R, Gurrapu S, Shivakrishna P. In silico identification of clinically approved medicines against the main protease of SARS-CoV-2, causative agent of covid-19. arXiv: Biomolecules 2020.

[10] UBC. Apeiron Biologics to Trial Coronavirus Drug Candidate in China. Available online: 2021.https://www.clinicaltrialsarena.com/news/ubc-apeiron-biologics-covid-19-trial/ (accessed on 31 May 2021).

[11] Belouzard S, Chu VC, Whittaker GR. Activation of the SARS coronavirus Spike protein via sequential proteolytic cleavage at two distinct sites. Proc. Natl. Acad. Sci. U S A. 2009;106:5871–5876

[12] Hoffmann M, Kleine Weber, H, Schroeder S, Krüger N, Herrler T, Erichsen S et al. SARS-CoV-2 cell entry depends on ACE2 and TMPRSS2 and is blocked by a clinically proven protease inhibitor. Cell 2020;181: 271.

[13] Ajoy Basak, Abdel-Majid Khatib, Dayani Mohottalage, Sarmistha Basak, Maria Kolajova, Subhendu Sekhar Bag, Amit Basak Basak A.. A Novel Enediynyl Peptide Inhibitor of Furin That Blocks Processing of proPDGF-A, B and proVEGF-C. PLOS ONE 2009;4:11: e7700.

[14] Lu R, Zhao X, Li J, et al. Genomic characterisation and epidemiology of 2019 novel coronavirus: implications for virus origins and receptor binding. Lancet (London, England) 2020;395:565–74.

[15] Wang C, Horby PW, Hayden FG, Gao GF. A novel coronavirus outbreak of global health concern. The Lancet 2020;395:470–473.

[16] Caly L, Druce J D, Catton M G, Jans D A, Wagstaff K M. The FDA-approved drug ivermectin inhibits the replication of SARS-CoV-2 *in vitro.* Antiviral Res 2020;178:104787.

[17] Wang M, Cao R, Zhang L, Yang X, Liu J, Xu M. Remdesivir and chloroquine effectively inhibit the recently emerged novel coronavirus (2019-nCoV) in vitro. Cell Res 2020;30:269–271.

[18] Mugisha C S, Vuong H R, Puray Chavez M, Kutluay S B. A facile Q-RT-PCR assay for monitoring SARS-CoV-2 growth in cell culture. bioRxiv 2020.

[19] Nugiel DA, Jacobs K, Kaltenbach RF, Worley T, Patel M, Meyer DT, et al. Preparation and structure-activity relationship of novel P1/P1'-substituted cyclic urea-based human immunodeficiency virus type-1 protease inhibitor. Journal of Medicinal Chemistry1996;39 (11): 2156–2169.

[20] Patel M, Bacheler LT, Rayner MM, Cordova BC, Klabe RM, Erickson Viitanen S, Seitz SP. The synthesis and evaluation of cyclic ureas as HIV protease inhibitors: modifications of the P1/P1' residues. Bioorganic & Medicinal Chemistry Letters1998;8(7):823-28.

[21] Accessed from: https://www.rcsb.org/

[22] Abraham M J, Murtola T, Schulz R, Páll S, Smith J C, Hess B. GROMACS: high performance molecular simulations through multilevel parallelism from laptops to supercomputers. SoftwareX 2015;1:19–25.

[23] Estari Mamidala, Rakesh Davella, Pujala Shivakrishna. Spermine phosphate Inhibits the SARS-CoV-2 Spike–ACE2 Protein-Protein Interaction–as an in silico approach contribute to its antiviral activity against COVID-19. Annals of the Romanian Society for Cell Biology 2021;4814–4827.

[24] Coutard B. The spike glycoprotein of the new coronavirus 2019-nCoV contains a furin-like cleavage site absent in CoV of the same clade. Antivir Res 2020;176:104742.

[25] Tang T, Bidon M, Jaimes J A, Whittaker G R, Daniel S. Coronavirus membrane fusion mechanism offers as a potential target for antiviral development. Antiviral Res 2020; 104792.

[26] Coutard B. The spike glycoprotein of the new coronavirus 2019-nCoV contains a furin-like cleavage site absent in CoV of the same clade. Antiviral Res 2020;176:104742.

[27] Wrapp D, Wang N, Corbett KS, Goldsmith JA, Hsieh CL, Abiona O. Cryo-EM structure of the 2019-nCoV spike in the prefusion conformation. Science 2020;367:1260–1263.

[28] Z Jin, Y Zhao, Y Sun, B Zhang, H Wang, Y Wu . Structural basis for the inhibition of SARS-CoV-2 main protease by antineoplastic drug carmofur. Nat. Struct. Mol. Bio l2020; 27:529–532.

[29] MTJ Quimque, KIR Notarte, RAT Fernandez, MAO Mendoza, RAD Liman, JAK Lim. Virtual screeningdriven drug discovery of SARS-CoV2 enzyme inhibitors targeting viral attachment, replication, post-translational modification and host immunity evasion infection mechanisms. J. Biomol. Struct. Dyn 2020:1–23.

[30] M U Mirza, M Froeyen. Structural elucidation of SARS-CoV-2 vital proteins: computational methods reveal potential drug candidates against main protease, Nsp12 polymerase and Nsp13 helicase. J. Pharm. Anal 2020;10(4):320-328.

[31] Z Jin, X Du, Y Xu, Y Deng, M Liu, Y Zhao, B Zhang. Peng, Structure of M pro from SARS-CoV-2 and discovery of its inhibitors. Nature 2020;1–5.

[32] A Shannon, NTT Le, B Selisko, C Eydoux, K Alvarez, JC Guillemot, E Decroly. Remdesivir and SARS-CoV-2: Structural requirements at both nsp12 RdRp and nsp14 Exonuclease active-sites. Antiviral Res 2020;104793.

[33] W Yin, C Mao, X Luan, DD Shen, Q Shen, H Su, X Wang, F Zhou, W Zhao, M Gao. Structural basis for inhibition of the RNA-dependent RNA polymerase from SARS-CoV-2 by remdesivir. Science 2020;368:1499-1504

[34] Y Wang, V Anirudhan, R Du, Q Cui, L Rong. RNA-dependent RNA polymerase of SARS-CoV-2 as a therapeutic target. J. Med. Virol 2020;26264.

[35] KA Ivanov, J Ziebuhr. Human coronavirus 229E nonstructural protein 13: characterization of duplex-unwinding, nucleoside triphosphatase, and RNA 5′ - triphosphatase activities. J. Virol 2004;78:7833-7838.

[36] S Habtemariam, S F Nabavi, M Banach, I Berindan Neagoe, K Sarkar, PC Sil, S M Nabavi. Should we try SARS-CoV-2 helicase inhibitors for COVID-19 therapy. Arch. Med. Res 2020;51(7):733–735.

[37] J Lan, J Ge, J Yu, S Shan, H Zhou, S Fan, Q Zhang, X Shi, Q Wang, L Zhang. Structure of the SARS-CoV-2 spike receptor-binding domain bound to the ACE2 receptor. Nature 2020;581:215-220

[38] Z Liu, X Xiao, X Wei, J Li, J Yang, H Tan, J Zhu, Q Zhang, J Wu, L Liu. Composition and divergence of coronavirus spike proteins and host ACE2 receptors predict potential intermediate hosts of SARS-CoV-2. J. Med. Virol 2020;92:595-601.

[39] Hoffmann M, Hofmann Winkler H, Pohlmann S. Priming Time: How Cellular Proteases Arm Coronavirus Spike Proteins. Activation of Viruses by Host Proteases 2018;71-98.

**Table 1. Binding energy (kcal/mol) and intermolecular interactions of mozenavir drug against various protein targets involved in SARS-CoV-2 infection by molecular docking study.**

| Sl No | Target proteins | Intermolecular interactions against SARS-CoV-2 target proteins | | | |
|---|---|---|---|---|---|
| | | Binding Energy (kcal/mol) | Number of Hydrogen bonds | Hydrogen bonding interactions | Hydrophobic interactions |
| 1 | Furin (5JXH) | -12.04 | 5 | TYR308, GLY265, GLY255, ASP154, VAL231 | GLU236, PRO256, TRP254, LEU227, HIS194, GLY229, ASP264, ASP191, GLU230, GLY229 |
| 2 | ACE-2 (7DF4) | -9.71 | 4 | ARG393, TYR385, HIS378, GLU375 | ASN394, ASP350, HIS401, ALA348, HIS374, PRO346, HIS345, PHE504, THR347, PHE40 |
| 3 | Mpro (6Y2E) | -8.79 | 2 | ASP153, ASP245 | ILE249, PHE294, HIS246, GLY109, ILE200, GLN110, THR292, ASN203, VAL202, THR111, ASN151 |
| 4 | RdRp (7B3C) | -7.32 | - | - | TYR515, TRP509, ALA375, LEU371, MET380, TYR374, PHE340, PHE368, PHE506, LEU372 |
| 5 | Spike protein (2AJF) | -7.09 | 2 | ASP415, SER370 | VAL369, GLY368, TYR367, ASP414, CYS366, LYS365, PRO399, GLN401 |
| 6 | TMPRSS2 (7MEQ) | -7.08 | 1 | TYR337 | ALA490, GLN487, PRO335, HIS334, ILE332, SER333, ASN303, VAL298, VAL331, LYS330 |

**Table 2. Binding energy in kcal/mol for the test drug mozenavir along with the reference drugs against each protein target of SARS-CoV-2**

| Sl.No | Name of the Drug | Binding energy (kcal/mol) against SARS-CoV-2 target proteins | | | | | |
|---|---|---|---|---|---|---|---|
| | | Furin | ACE2 | Mpro | RdRp | Spike | TMPRSS2 |
| 1 | Mozenavir | -12.04 | -9.71 | -8.79 | -7.32 | -7.09 | -7.08 |
| Ref | Decanoyl-RVKR-chloromethylketone | -6.89 | | | | | |
| Ref | Chloroquine phosphate | | -7.88 | | | | |
| Ref | Indinavir | | | -7.11 | | | |
| Ref | Remdesivir | | | | -4.7 | | |
| Ref | Arbidol | | | | | -7.86 | |
| Ref | Camostat | | | | | | -5.9 |

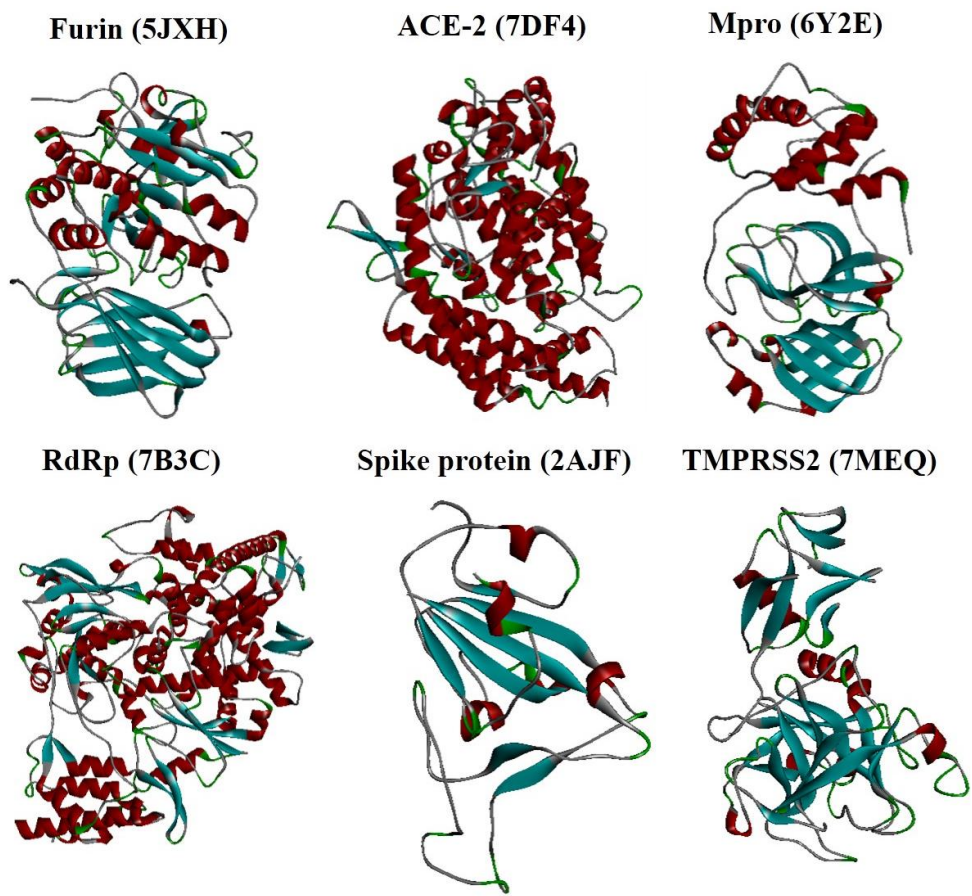

**Figure 1. 3-Dimentional structures of SARS-CoV-2 target proteins**

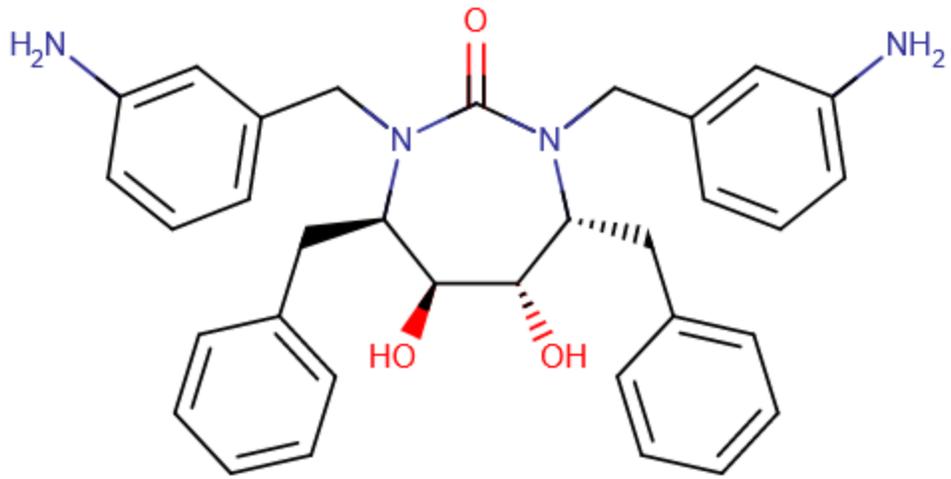

**Figure 2. 2-Dimentional structure of Mozenavir drug**

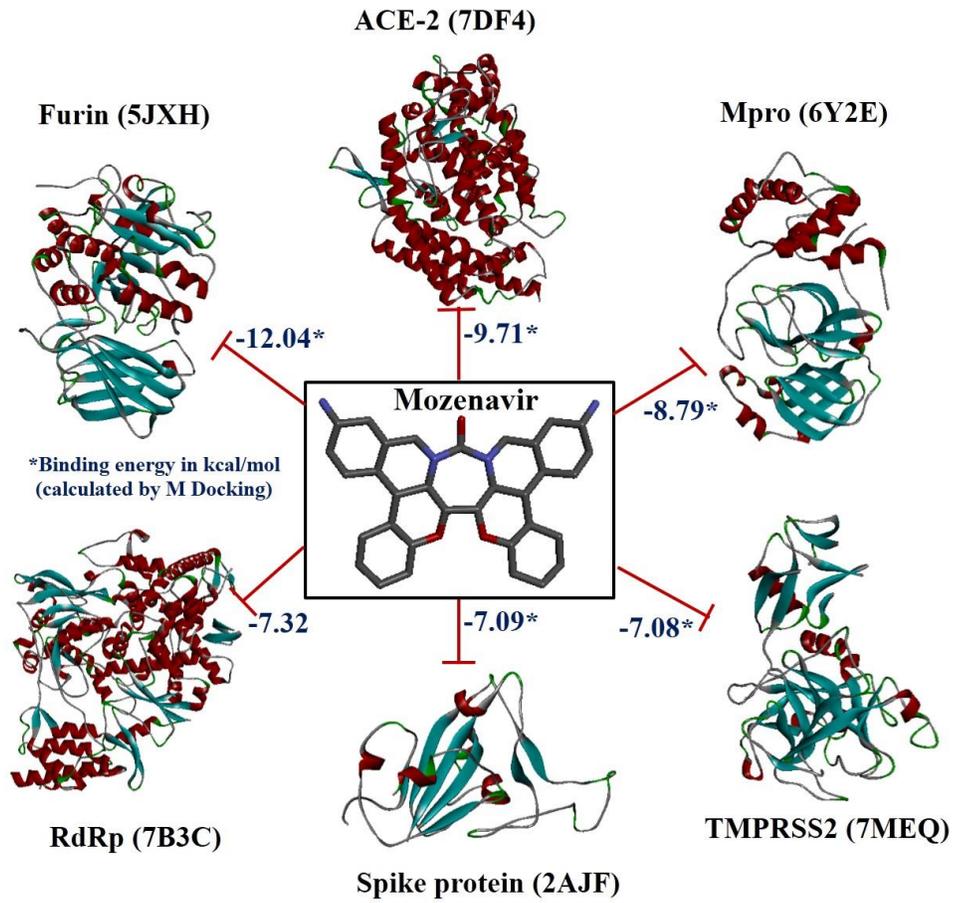

**Figure 3. Illustration of mozenavir inhibition against the SARS-CoV-2 target proteins: Furin, ACE2, Mpro, RdRp, Spike Glycoprotein, and TMPRSS2**

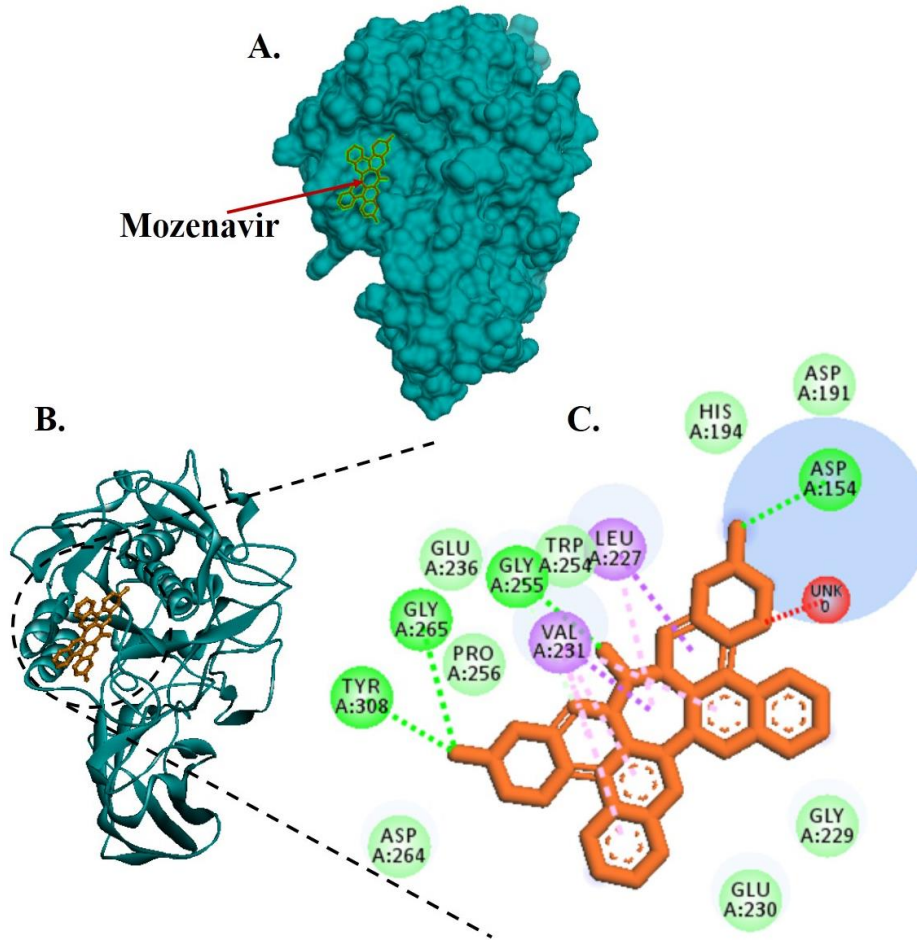

**Figure 4. Docking interactions of Mozenavir with furin protease (PDB ID: 5JXH)**

(a) Surface structure of best binding mode in the protein pocket (ligand illustrated as orange sticks), (b) 3D structure of amino acid residues involved in the interaction with mozenavir ligand (ligand as orange sticks), and (c) 2D structure of mozenavir binding interaction with amino acid with a hydrogen bond (green dashed line).

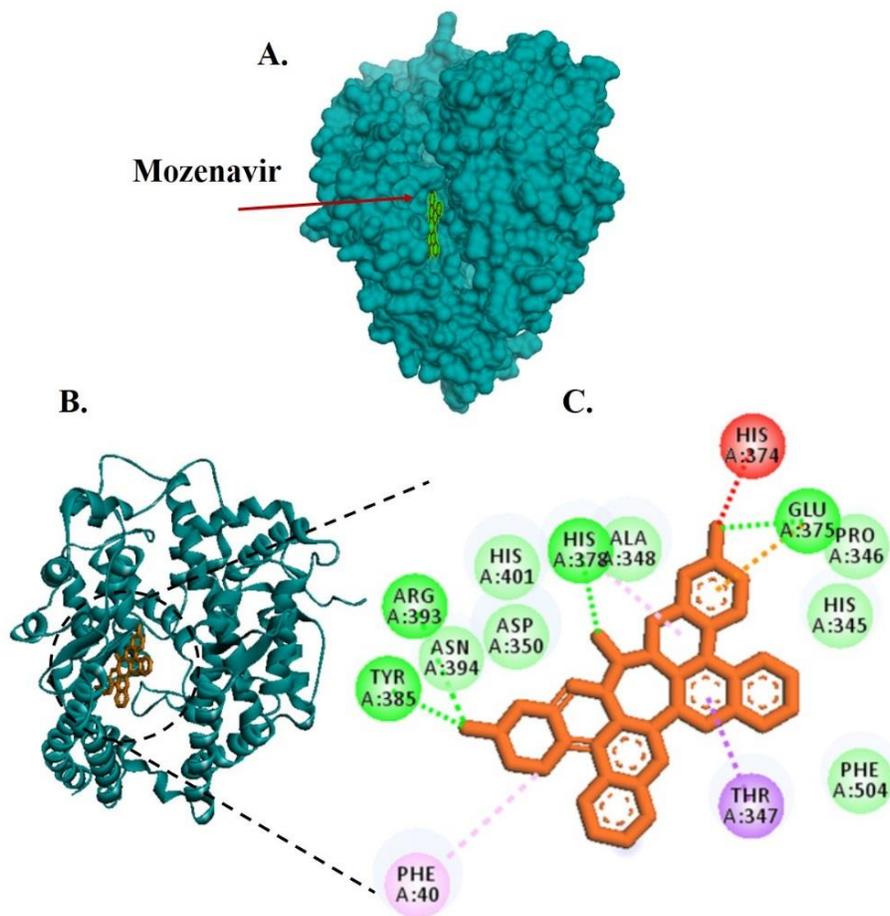

**Figure 5. Docking interactions of Mozenavir with ACE-2 (PDB ID: 7DF4)**

(a) Surface structure of best binding mode in the protein pocket (ligand illustrated as orange sticks), (b) 3D structure of amino acid residues involved in the interaction with mozenavir ligand (ligand as orange sticks), and (c) 2D structure of mozenavir binding interaction with amino acid with a hydrogen bond (green dashed line).

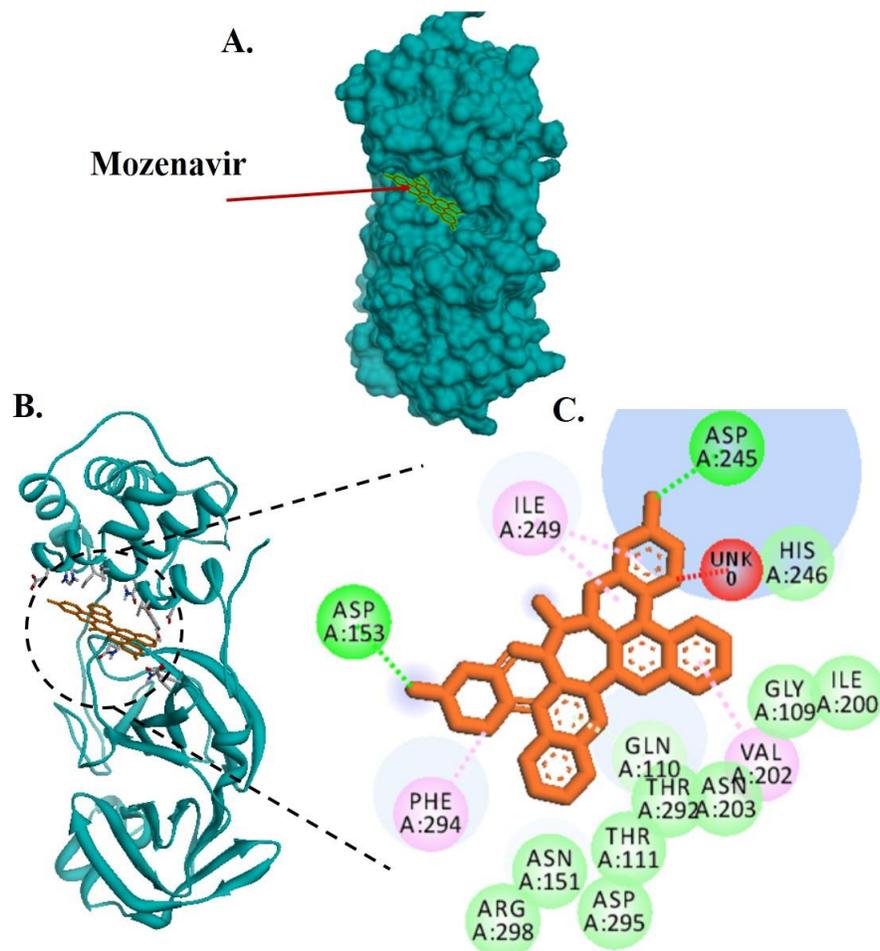

**Figure 6. Docking interactions of Mozenavir with the SARS Main protease (Mpro) (PDB ID: 6YTE)**

(a) Surface structure of best binding mode in the protein pocket (ligand illustrated as orange sticks), (b) 3D structure of amino acid residues involved in the interaction with mozenavir ligand (ligand as orange sticks), and (c) 2D structure of mozenavir binding interaction with amino acid with a hydrogen bond (green dashed line).

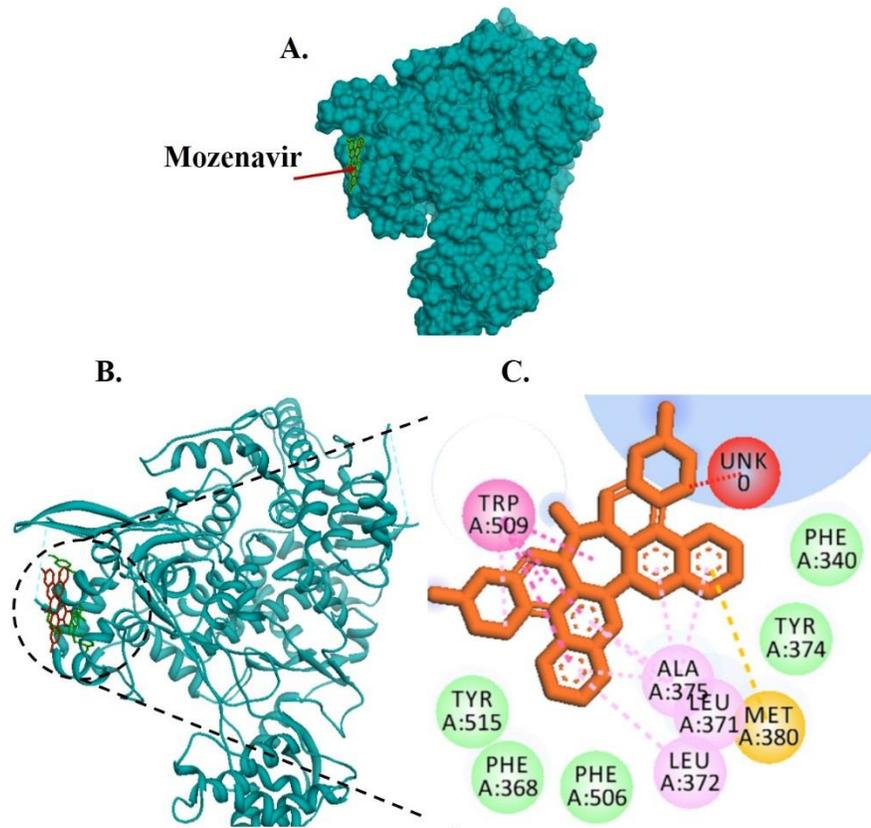

**Figure 7. Docking interactions of Mozenavir with the SARS-CoV-2 RdRp (PDB ID: 7B3C)**

(a) Surface structure of best binding mode in the protein pocket (ligand illustrated as orange sticks), (b) 3D structure of amino acid residues involved in the interaction with mozenavir ligand (ligand as orange sticks), and (c) 2D structure of mozenavir binding interaction with amino acid with a hydrogen bond (green dashed line).

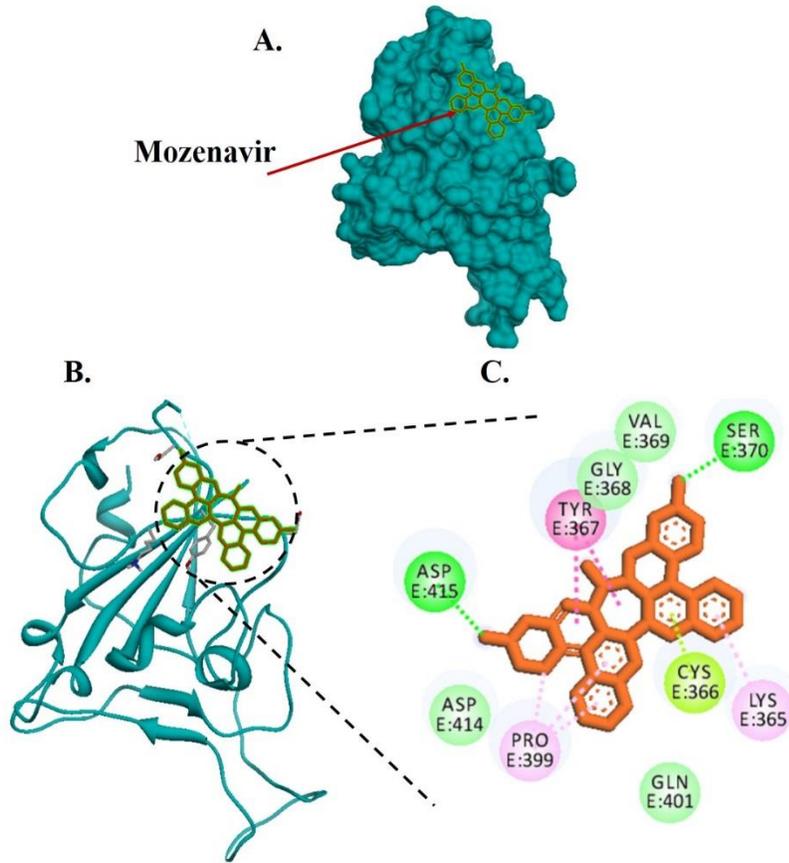

**Figure 8. Docking interactions of Mozenavir with the SARS-CoV-2 spike glycoprotein (PDB ID: 2AJF)**

(a) Surface structure of best binding mode in the protein pocket (ligand illustrated as orange sticks), (b) 3D structure of amino acid residues involved in the interaction with mozenavir ligand (ligand as orange sticks), and (c) 2D structure of mozenavir binding interaction with amino acid with a hydrogen bond (green dashed line).

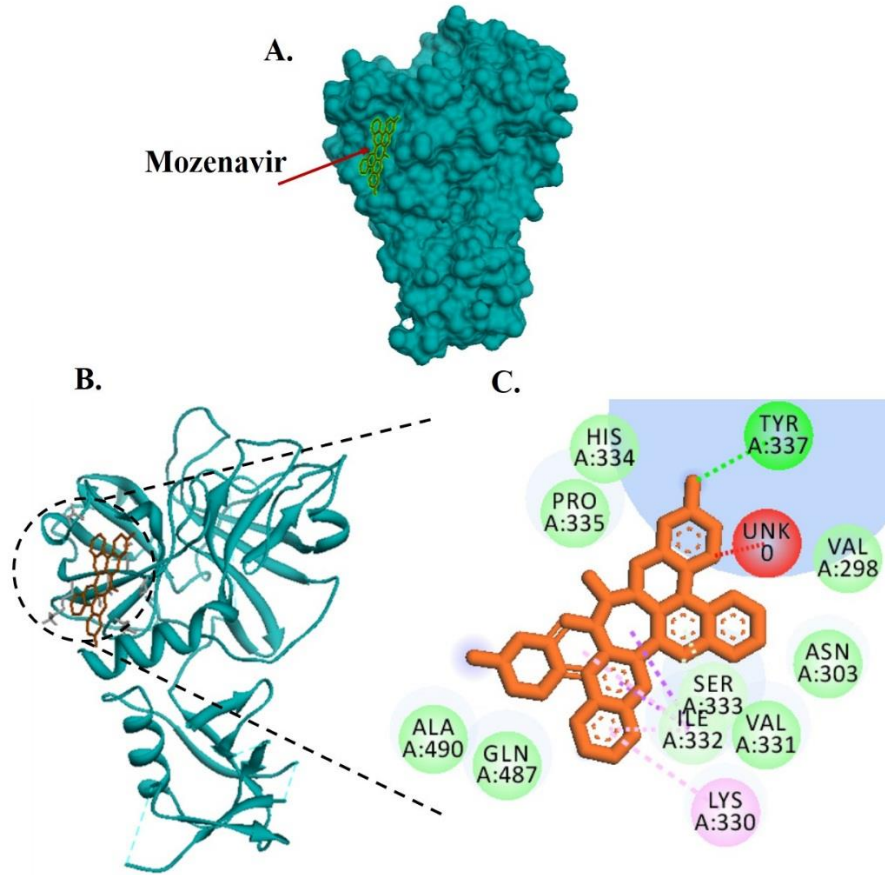

**Figure 9. Docking interactions of Mozenavir with TMPRSS2 (PDB ID: 7MEQ)**

(a) Surface structure of best binding mode in the protein pocket (ligand illustrated as orange sticks), (b) 3D structure of amino acid residues involved in the interaction with mozenavir ligand (ligand as orange sticks), and (c) 2D structure of mozenavir binding interaction with amino acid with a hydrogen bond (green dashed line).

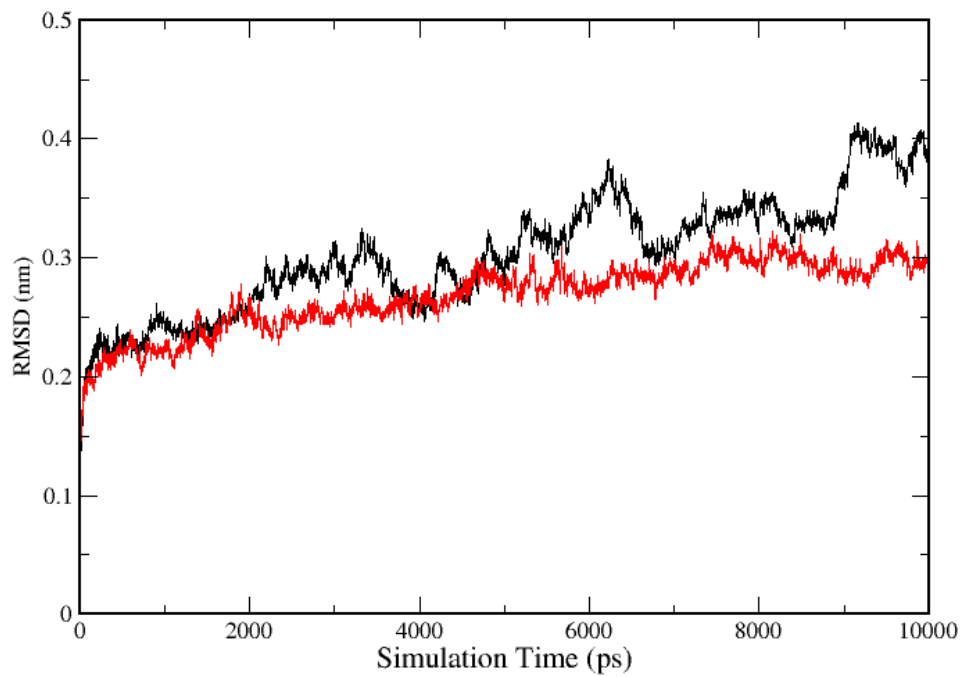

**Figure 10.** Plot of root mean square deviation (RMSD) during 10ns MD simulation of SARS-CoV-2 target protein furin alone (black colour) in complex with mozenavir (red colour)

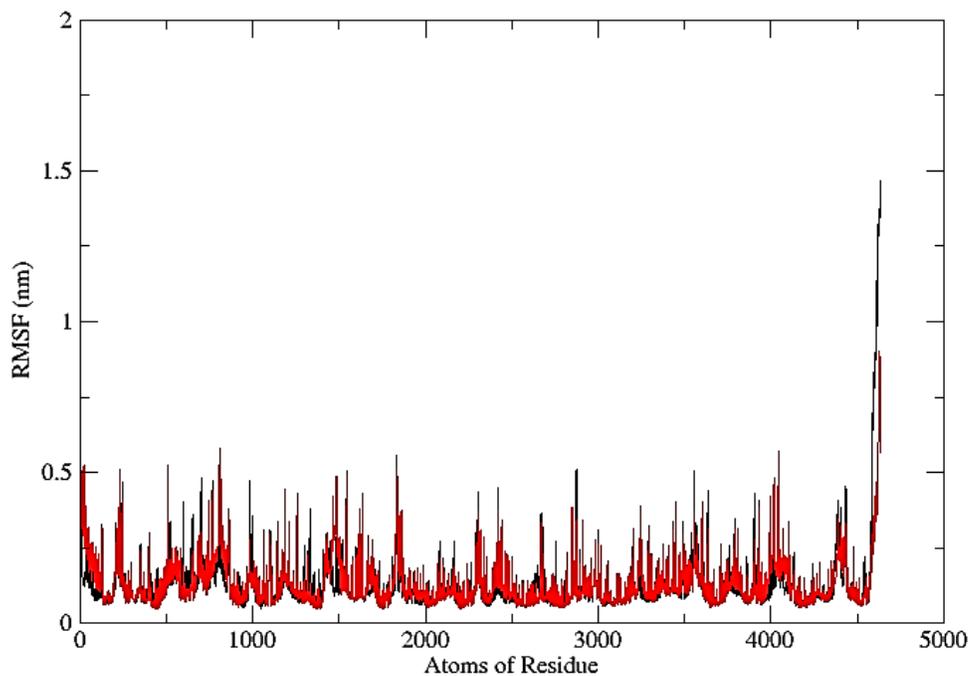

**Figure 11. Plot of root mean square fluctuation (RMSF) values, during 10ns MD simulation of SARS CoV-2 target protein furin alone (black colour) and in complex with mezanovir (red colour)**